\documentclass[npg]{copernicus}
\begin{document}

\title{An experimental study of regime transitions in a differentially heated baroclinic annulus with flat and sloping bottom topographies}
\author[1]{M. Vincze}
\author[1]{U. Harlander}
\author[2]{Th. von Larcher}
\author[1]{C. Egbers}

\affil[1]{Department of Aerodynamics and Fluid Mechanics, Brandenburg University of Technology Cottbus-Senftenberg, Cottbus, Germany}
\affil[2]{Institute for Mathematics, Freie Universit\"at Berlin, Berlin, Germany}
\runningtitle{Regime transitions in a baroclinic annulus with flat and sloping bottom topographies}
\runningauthor{M. Vincze et al.}
\correspondence{M. Vincze\\ (vincze@tu-cottbus.de)}
\received{}
\pubdiscuss{} 
\revised{}
\accepted{}
\published{}

\maketitle
\begin{abstract}
A series of laboratory experiments has been carried out in a thermally driven rotating annulus to study
the onset of baroclinic instability, using horizontal and uniformly sloping bottom topographies. Different wave flow regimes have been identified and their phase boundaries -- expressed in terms of appropriate non-dimensional parameters -- have been compared to the recent numerical results of \citet{thomas_slope}. In the flat bottom case, the numerically predicted alignment of the boundary between the axisymmetric and the regular wave flow regime was found to be consistent with the experimental results.
However, once the sloping bottom end wall was introduced, the detected behaviour was qualitatively different from that of the simulations. This disagreement is thought to be the consequence of nonlinear wave-wave interactions that could not be resolved in the framework of the numerical study. This argument is supported by the observed development of interference vacillation in the runs with sloping bottom, a mixed flow state in which baroclinic wave modes exhibiting different drift rates and amplitudes can co-exist.       
\end{abstract}    

\section{Introduction}
Based on the principle of hydrodynamical similarity, some fundamental characteristics of Earth's various large-scale atmospheric flow phenomena can be modelled using surprisingly simple tabletop-size experimental set-ups. Under laboratory conditions it is possible to control the governing physical parameters and thus to separate different processes that cannot be studied independently in such a complex system as the real atmosphere. Therefore, laboratory experiments provide a remarkable test bed to validate numerical techniques and models aiming to investigate flows observed in the atmosphere and in the oceans.     

A classic apparatus to demonstrate the basic large-scale dynamics of the mid-latitude atmosphere
is the differentially heated rotating annulus, introduced by \citet{fultz_alap} and \citet{hide_alap}, based on the principles suggested by Vettin in the mid-19th century \citep{vettin}. 
The set-up (Fig.\ref{setup}) consists of a cylindrical gap rotating around its vertical axis of symmetry, with cooled inner side wall and a heated outer side wall, thus the working
fluid (usually water or silicone oil) experiences a radial temperature gradient. These boundary conditions mimic the meridional differential incoming solar heat flux on Earth. The characteristic hydrodynamical timescale scales with the temperature difference $\Delta T$ between the inner and the outer side wall, and without rotation, the velocity field would be axisymmetric. (Note, that in such a `sideways convection' configuration, where the heating and cooling takes place at \emph{vertical} boundaries, there is no minimum $\Delta T$, i.e. \emph{any} finite temperature difference would initiate an overturning flow.) However -- since rotation is present -- in the co-rotating reference frame Coriolis' force also acts on the flow, with a magnitude proportional to the flow velocity and the rotation rate $\Omega$.

The ratio of the revolution period and the aforementioned hydrodynamical timescale yields an appropriate non-dimensional number for temperature-driven rotating systems, the thermal Rossby number $Ro_T$ (see Section 2). For a fluid like water, in case of $Ro_T \gg 1$, the flow is axisymmetric and not significantly disturbed by rotation, whereas for $Ro_T \ll 1$ (as in the case of cyclones and and anticyclones in the atmosphere) the dynamics is dominated by the Coriolis effect. Another important non-dimensional parameter, the Taylor number, $Ta$, can be obtained similarly, by measuring the timescale of viscous effects with respect to the rotation period (more details in Section 2). $Ro_T$ and $Ta$ are used to characterise the different dynamical regimes in the rotating annulus. 

Between the axisymmetric and geostrophic turbulent flow state at large Taylor numbers there is a certain region on the $Ta-Ro_T$ regime diagram, where the velocity and temperature fields exhibit persistent regular wave-like patterns that propagate along the azimuthal direction in the tank due to \emph{baroclinic instability} (see Fig.\ref{thomas_replot}, or for a more general qualitative view Fig.1 of \citet{thomas_npg}). The boundary separating the axisymmetric and unstable states on the regime diagram (e.g. the blue and green curves in Fig.\ref{thomas_replot}) is referred to as the `transition curve'.

The basic underlying physics of such baroclinic waves has been subject of extensive theoretical \citep[e.g.][]{eady, mason, lorenz}, numerical \citep[e.g.][]{torsten, read} and experimental \citep[e.g.][]{frueh,thomas_npg,uwe_obst} research throughout the past decades.
Furthermore, some studies focused on the quantitative comparison of temperature statistics \citep{gyure} and propagation dynamics of passive tracers \citep{viki} obtained from annulus experiments and from actual atmospheric data.   

The different types of \emph{vacillation} phenomena (termed `amplitude', `structural' and `interference' vacillation) have attracted considerable attention within the community of geophysical fluid dynamics and of nonlinear systems, dating back to the classic work of \citet{lorenz}. The co-existence of different baroclinic wave modes and their interactions in a rotating annulus (with flat bottom) has been the scope of experimental studies, e.g. by \citet{pfeffer} or \citet{uwe_piv}.    

As another example of simple laboratory-scale experiments of geophysical relevance, rotating cylindrical tanks with conical bottom end wall are broadly used as a minimal model of the concept of $\beta$-effect, and in particular to demonstrate topographic (barotropic) Rossby waves, driven by conservation of potential vorticity over a sloping bottom \citep{rossby}. 
However, the wave flow phenomena in a combined differentially heated \emph{and} sloping set-up have been surprisingly rarely investigated experimentally.
The first of the very few studies addressing this issue was the pioneering work of \citet{fultz_kaynor}, who first reported wave dispersion of baroclinic waves in the presence of tilted bottom topography, and tried to quantify the group velocities of the propagating patterns and compare their results to the known dispersion relation of Rossby waves. 
A different approach was followed by \citet{mason}, who applied a modified version of Eady's aforementioned theoretical framework to interpret their experimental results on the effect of different sloping top and bottom end wall configurations on the flow stability.    
 
A recent linear numerical study by \citet{thomas_slope} investigated how the presence of a sloping bottom end wall alters the wave flow regime boundaries, applying the exact geometrical dimensions of the baroclinic wave tank used for experimental works at the Brandenburg Technical University since about 2003, cf., \citet{thomas_npg}. The linear stability analysis of that work revealed significant differences between the flat and sloping bottom end wall configurations. While the overall typical anvil shape of the wave flow regime on the $Ta-Ro_T$ plane remains similar in the sloping bottom case, the transition curve (more precisely, the neutral linear stability curve) has shifted significantly towards lower values of $Ta$, and at higher values of $Ro_T$ a sharp reversal of the curve has been found (Fig.\ref{thomas_replot}). The comparison of these numerical results to the behaviour of the actual experiment was the main focus and primary motivation of the present work. We also aimed to obtain a more detailed picture on wave-wave interactions and to gain a deeper insight to the structure of the transition region between the axisymmetric and wave flow regimes.   

Our paper is organized as follows. Section 2 outlines the experimental set-up, the parameters that describe the state of the system and the measurement techniques used. Section 3 discusses the applied data evaluation methods. The results are presented in the four subsections of Section 4. In Section 5 we summarize the results and discuss their implications on the physics of the underlying dynamics.   
 
\section{Experimental apparatus and methods}
\begin{figure}[h!]
\noindent\includegraphics[width=8.6cm]{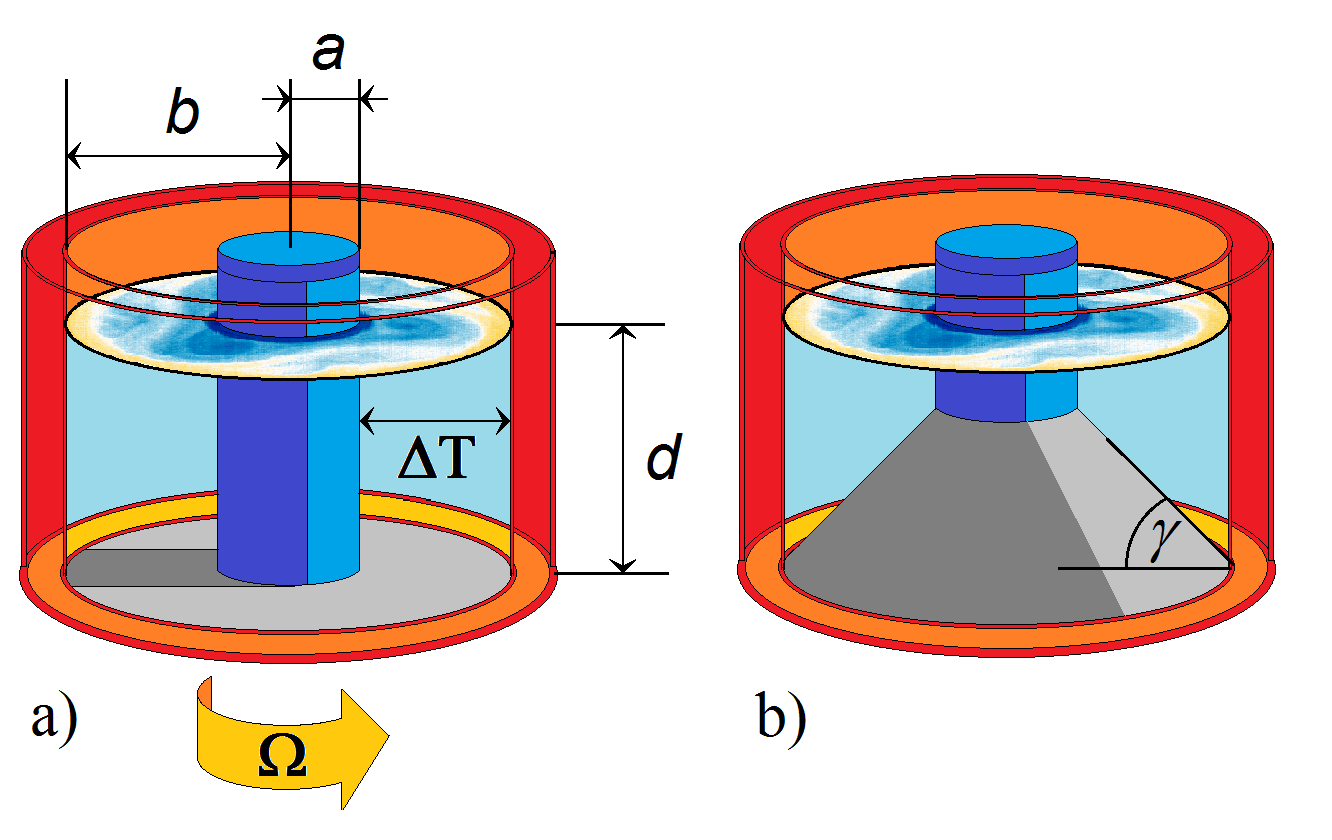}
 \centering
\caption{Schematic drawing of the set-up with a flat (a) and with a sloping bottom end-wall. For the geometric parameters indicated, see text. The counter-clockwise direction of rotation is indicated.}
\label{setup}
\end{figure}

Our experiments have been carried out in a circular laboratory tank mounted on a turntable. The tank consisted of three concentric cylinders, rotating at the same rate in counter-clockwise direction (at angular velocities between $\Omega=0.13-0.83$ rad/s), whose geometric center coincided with the axis of rotation, as depicted in Fig.\ref{setup}.
In the inner cylinder, made of anodized aluminium, cold water was circulated, the temperature of which was monitored by two thermocouples and adjusted via a cooling thermostat. The `outer ring' of the tank, i.e. the annular gap between the two outermost cylinders (painted red and orange in Fig.\ref{setup}), made of borosilicate glass, hosted the heating wire and contained water as heat conductive medium. 
The temperature difference $\Delta T$ between the outer ring and the inner cylinder has been varied between 2.4 and 8.0 K for the different experimental runs. The methods and characteristics of the temperature control have been discussed in more detail by \citet{thomas_npg}.  

The middle cavity of the annulus, ranging from $a=4.5$ cm to $b=12$ cm in radial direction (Fig.\ref{setup}a), was filled up with the working fluid -- de-ionized water -- to the height of $d = 13.5$ cm, yielding radius ratio $\eta=a/b=0.38$ and aspect ratio $\Gamma=d/(b-a)=1.8$. The surface of this cylindrical gap was \emph{free}, which, besides of its importance as boundary condition, also made it possible to use infrared thermography as our measurement technique (infrared radiation is generally absorbed by glass or acryllic, therefore thermography cannot be applied for set-ups with rigid top). 

The infrared camera was mounted above the middle of the wave tank and was co-rotating with the set-up. In every $\Delta t = 10$ s, $626\times 428$-pixel thermograms were taken, covering the surface of the annulus with a resolution of $\sim 0.03$ K. 
The patterns in these thermograms can be considered \emph{surface temperature} structures, since the penetration depth of the applied wavelength range into water is measured in millimetres. 

The most important dynamic control parameters, $\Omega$ and $\Delta T$, can be expressed in terms of non-dimensional numbers.
The former is captured by the Taylor number $Ta$, defined as:
\begin{equation}
Ta = \frac{4\Omega^2(b-a)^5}{\nu^2 d},
\label{Ta}
\end{equation} 
where $\nu=1.004\times 10^{-6}$ m$^2$/s is the kinematic viscosity of water. The ratio of the characteristic velocity of the thermally driven flow to the rotation rate yields the thermal Rossby number $Ro_T$, which serves as the non-dimensional $\Delta T$-scale:
\begin{equation}
Ro_T = \frac{g d \alpha \Delta T}{\Omega^2 (b-a)^2},
\label{Ro}
\end{equation}
where $\alpha=2.07\cdot10^{-4}$ K$^{-1}$ is the volumetric thermal expansion coefficient of freshwater and $g$ represents the acceleration due to gravity. The physical properties of the working fluid are captured by the Prandl number $Pr$:
\begin{equation}
Pr=\frac{\nu}{\kappa}=7.0,
\end{equation}
expressing the ratio of the kinematic viscosity $\nu$ and thermal conductivity $\kappa=0.1434\times 10^{-6}$ m$^2$/s of water. 

The experiments were conducted the following way: after setting the values of the first target rotation rate and the temperature difference, it took typically around 1.5--3 hours for the temperature control system to reach and stably maintain the required $\Delta T$. The thermographic measurements started when quasi-stationarity of the temperature signals was reached, and lasted for 2000--4000 s, that led to 200-400 snapshots per parameter point. After the first measurement was performed, $Ta$ was gradually increased or decreased (keeping $\Delta T$ constant) over a period of 1000--2000 s. When the next parameter point was reached, the thermographic measurement started immediately. Thus, the transient flow phenomena could also be recorded, before the system settled its next quasi-equilibrium state. The typical duration scale of this transient phase was 500--1000 s. In one measurement session (typically lasting for 6 hours) a maximum of 4 different $Ta - Ro_T$ parameter pairs could be investigated.           

The first series of experiments was conducted with flat bottom topography (Fig.\ref{setup}a), whereas in the second series, a conical sloping bottom obstacle made of polyamid was placed in the cylindrical gap of the tank (Fig.\ref{setup}b). The slope had an angle of $\gamma=35^\circ$ to the horizontal as in the paper by \citet{thomas_slope}, and was decreasing towards the outer side wall, so that the maximum depth $d$ (reached at distance $b$ from the axis) coincided with that in the flat bottom configuration.

\section{Data processing}
In this section we describe the techniques of data evaluation and demonstrate them exemplarily for the reader's convenience using a typical experimental run (with flat bottom end wall) where the flow was dominated by a regular wave pattern, exhibiting three-fold rotational symmetry.

\begin{figure}[h!]
\noindent\includegraphics[width=8.6cm]{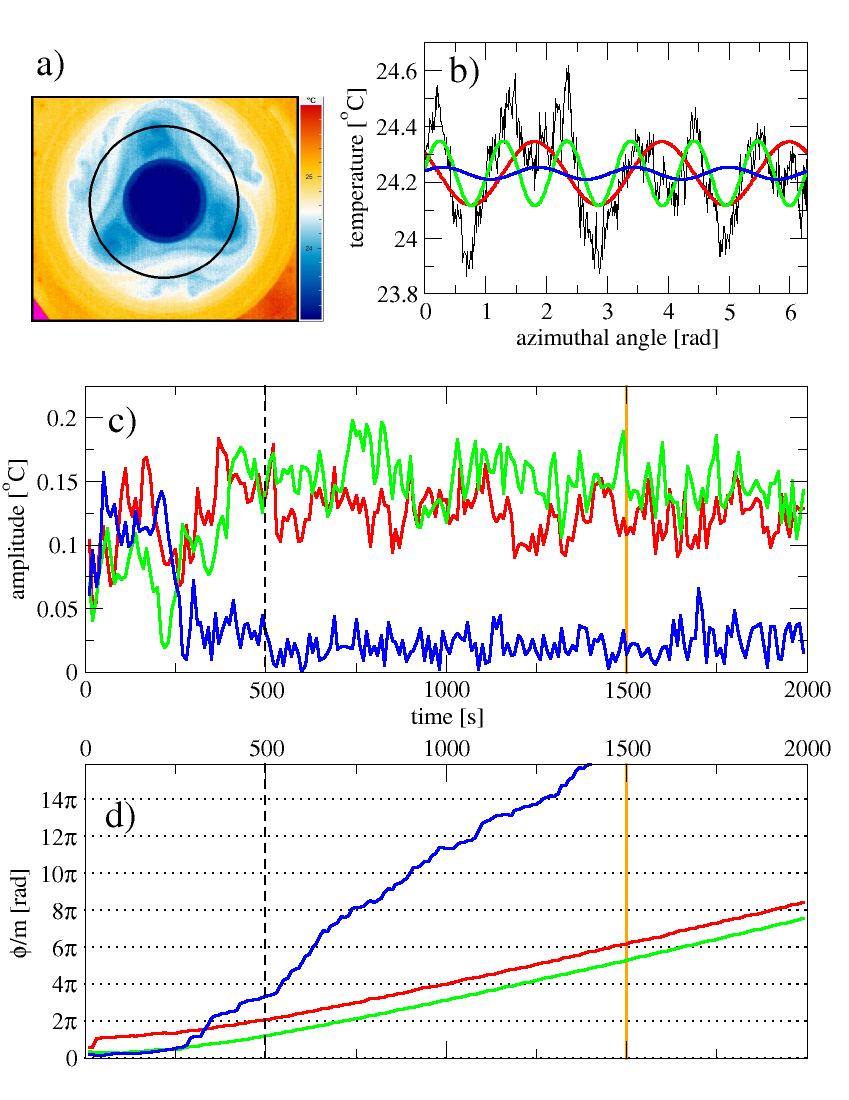}
 \centering
\caption{The steps of data processing. The temperature values from a `raw' thermographic image (a) are extracted along a circular contour (black circle in panel a) to obtain one-dimensional temperature data (b) for spatial Fourier transform. Modes $m=3,4$ and $6$ are shown by red, blue and green graphs, respectively. The temporal development of the ampliudes (c) and `azimuthal distances' (d) of these spectral components are shown using the same color coding.}
\label{fourierdemo}
\end{figure}

In order to reduce the investigated parameter space, a path-wise temperature series was extracted from each thermographic image (as the one seen in Fig.\ref{fourierdemo}a). The one-dimensional temperature data $T(\theta ; t)$ were obtained along a circular contour at mid-radius $r_{\rm mid}\equiv (a+b)/2$ (black circle in Fig.\ref{fourierdemo}a), as a function of the azimuthal angle $\theta$, measured clockwise from the `uppermost' point of the circle. The resulting data from this particular snapshot is shown by the black curve of Fig.\ref{fourierdemo}b.

For the analysis of the azimuthal temperature fields discrete spatial Fourier transform was used. After subtracting 
the mean temperature $\langle T(\theta ; t)\rangle$ (averaged over the whole $\theta=[0;2\pi]$ range at each time instant $t$), the remaining fluctuations could be decomposed into amplitudes $A_m(t)$ and phases $\phi_m(t)$ of normal modes of quantized wave numbers $m=1,...,6$, as:
\begin{equation}
T(\theta ; t)-\langle T(\theta ; t)\rangle \approx \sum_{m=1}^{6}A_m(t)\cdot \sin(m \theta + \phi_m(t)).
\label{decomp}
\end{equation} 
Fig.\ref{fourierdemo}b demonstrates this step, showing three (exemplarily selected) components: $m=3$ (red), $m=4$ (blue) and $m=6$ (green).
The temporal development of the amplitudes of these three modes is presented in Fig.\ref{fourierdemo}c, using the same color encoding. It is clearly visible, that after an initial transient phase, mode $m=4$ has decayed markedly, and from then on the wave pattern was dominated by the leading $m=3$ and $m=6$ modes. After the $t=500$ s mark (vertical dashed line in Figs.\ref{fourierdemo}c and d), the system stayed in a quasi-stationary state. (The time instant which corresponds to the patterns of Figs.\ref{fourierdemo}a and b is marked by an orange vertical line.)   

Panel d) shows the time series $\phi_m(t)/m$ for the above three modes, a quantity that measures the `azimuthal distance' travelled by the given component pattern since $t=0$. For better visualization, we extended the periodical $[0;2\pi]$ range to $[0;+\infty)$ (so that the positive increments correspond to counter-clockwise propagation). This also makes it easier to obtain values for the angular velocities $c_m(t)$ of the different modes, since:
\begin{equation}
\frac{1}{m}\frac{\partial \phi_m(t)}{\partial t}\equiv c_m(t),
\label{omega}
\end{equation}    
i.e. the values of $c_m(t)$ are the slopes of the curves in Fig.\ref{fourierdemo}d.
Visibly, modes $m=3$ and $m=6$ propagated at the same rate. 

Looking at panels a) and b), it is apparent that $m=3$ is the signature of the general three-fold symmetry of the wave pattern, whereas the presence of harmonic $m=6$ is related to the `fine structure' of the baroclinic wave, most notably to the cold vortices within the three warmer-than-average `lobes'.  Compared to these, the `azimuthal distance' of $m=4$ exhibited irregular behaviour after its amplitude decayed. In this case, the propagation of $m=4$ was labelled `non-physical'. This curve is presented here as an example to demonstrate that such signals could be easily distinguished from the clearly regular drift of the other two modes. This irregular behaviour can be thought of as a `random drift' of the phase $\phi_m(t)$. The small amplitude of this component (see Fig.\ref{fourierdemo}c) implies this being `noise' in the Fourier spectra instead of a clear physical signal.

Besides spatial Fourier transforms, standard tools of nonlinear time series analysis have also been applied, some details of which will be briefly discussed in Subsection 4.4. 

\section{Results}

\begin{figure}[t!]
\noindent\includegraphics[width=8.6cm]{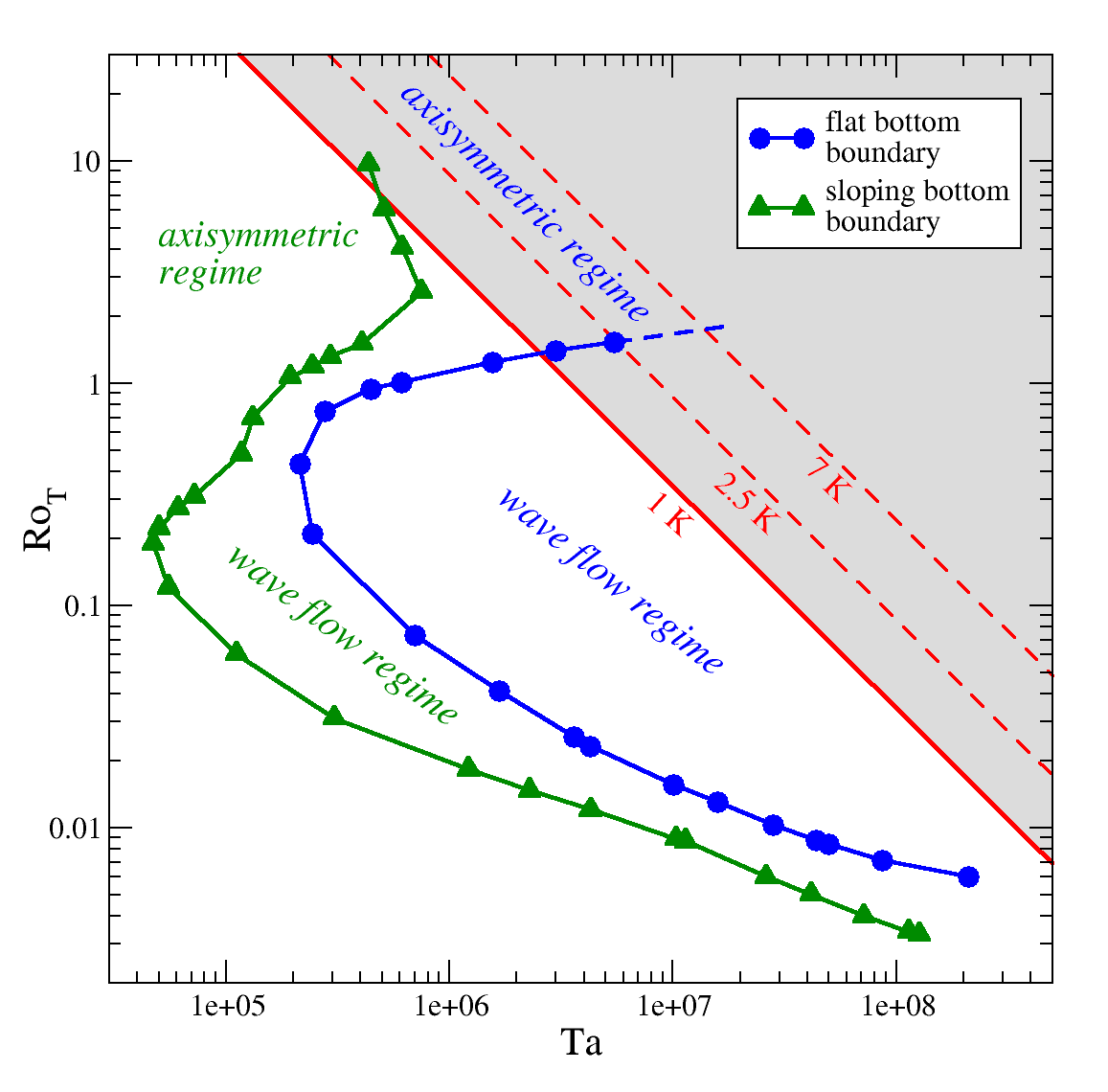}
 \centering
\caption{The neutral stability curves obtained by \citet{thomas_slope} for flat (blue) and sloping (green) bottom end walls.
The isotherms followed by our measurements are also indicated (dashed lines), as well as the parameter region that can be accessed with our techniques (grey shaded area).}
\label{thomas_replot}
\end{figure} 

Since the present study focused mainly on the transitions between the axisymmetric and the regular wave flow regimes, our experiments were confined to a certain domain of the $Ta$--$Ro_T$ parameter space in which such transitions were expected to occur according to the numerical results of \citet{thomas_slope}.
Their curves of neutral linear stability are depicted in Fig.\ref{thomas_replot} (both for the flat and sloping bottom case). The shaded area denotes the subset of $Ta$--$Ro_T$ parameter pairs, that could actually be reached and steadily maintained in our setup. This constraint is due to the fact that for temperature differences $\Delta T<1$ K the thermal fluctuations coming from technical constraints are almost comparable to the magnitude of $\Delta T$. 
Therefore the overlap between the region covered by the numerical study and the one within the reach of our methods is rather small. 

With both geometric configurations the experiments were carried out in the vicinity of the $\Delta T=2.5$ K and $7$ K isotherms in a domain ranging between $Ta=(0.124-4.90)\times 10^{7}$ and $Ro=0.55-16.42$.
Note, that -- as visible in Fig.\ref{thomas_replot} -- the $\Delta T=7$ K curve already falls out of the range of the numerical runs; here we applied linear extrapolation (dashed blue line in Fig.\ref{thomas_replot}) to approximate the expected regime boundary. 
    
\subsection{Experiments with flat bottom endwall}
\begin{figure}[b!]
\noindent\includegraphics[width=8.6cm]{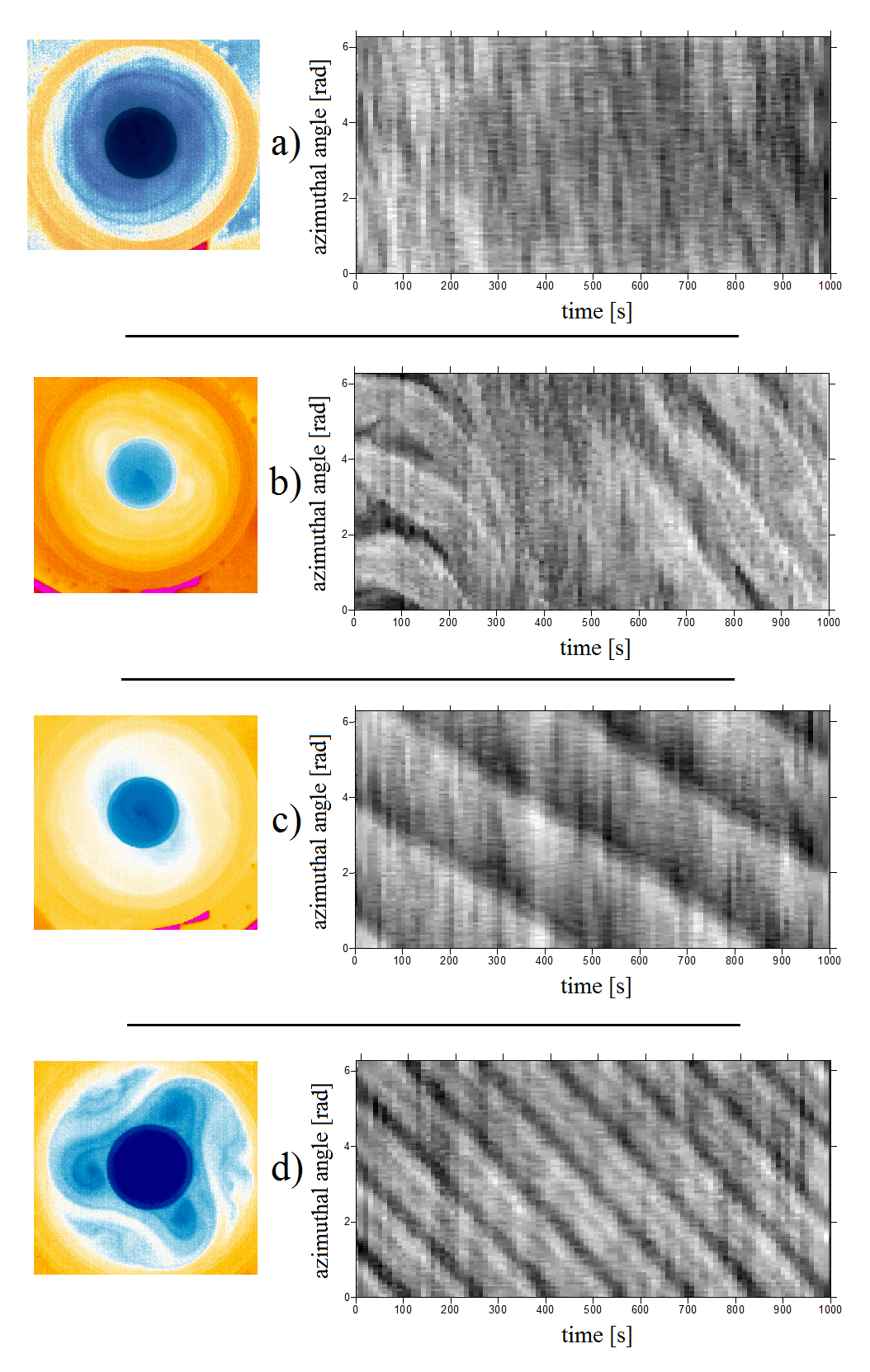}
 \centering
\caption{Thermographic snapshots (left) and Hovm\"oller plots (right) of some experimental runs with flat bottom topography. 
a) Axisymmetric ($m=0$) basic state ($Ro_T=2.73$; $Ta=3.31\times 10^6$), b) dispersive fluctuating weak wave (WW) state with traces of modes $m=2,3$ ($Ro_T=2.43$; $Ta=3.60\times 10^6$), c) regular wave with $m=2$ being the dominant component ($Ro_T=1.98$; $Ta=4.59\times 10^6$), d) regular pattern with leading component $m = 3$ ($Ro_T=1.55$; $Ta=5.78\times 10^6$).}
\label{hm_flat}
\end{figure}
\begin{figure}[h!]
\noindent\includegraphics[width=8.6cm]{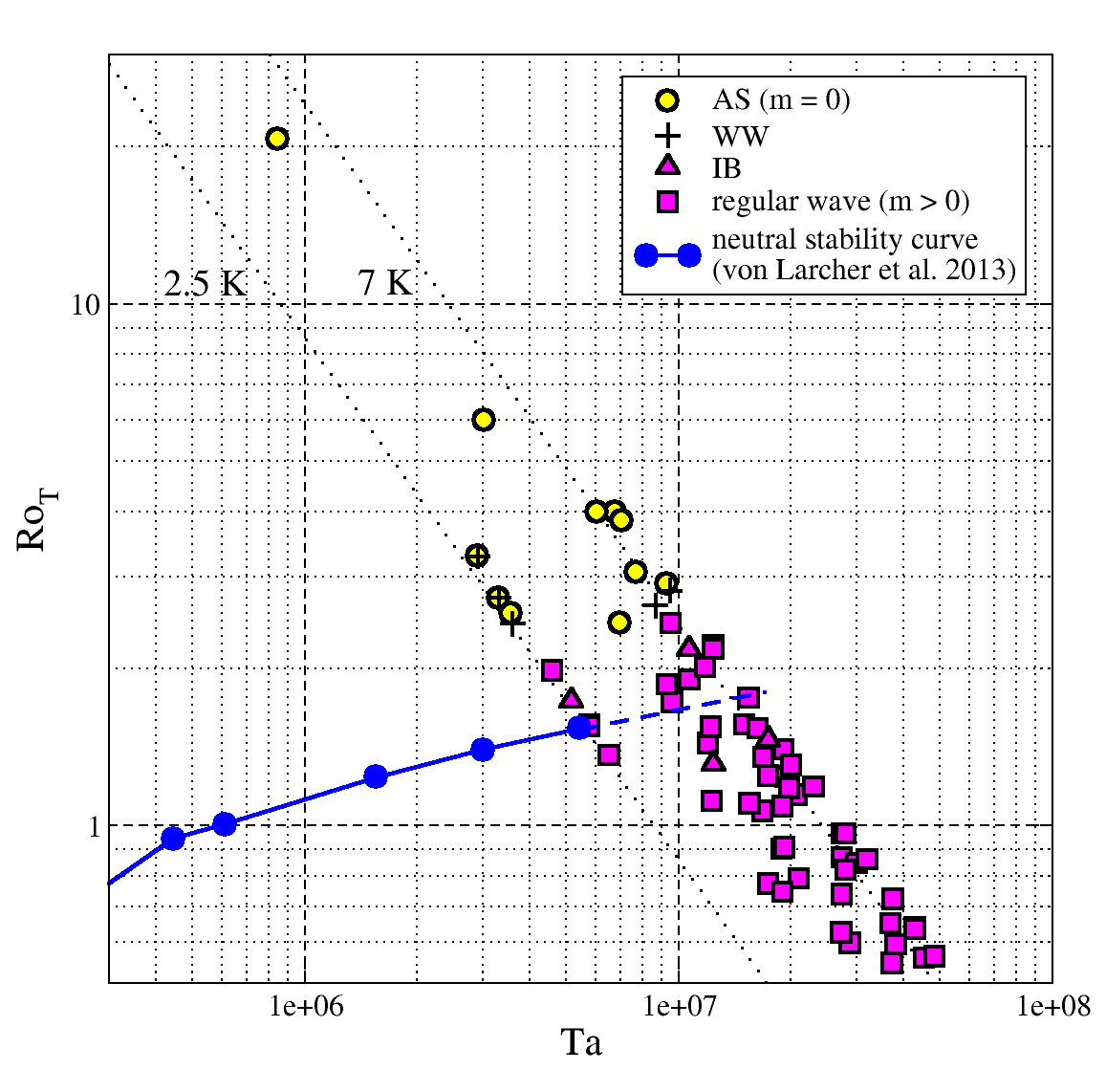}
 \centering
\caption{The regime diagram for the flat bottom experiments. The neutral numerical stability curve is marked by the blue solid curve (extended as a dashed line for the region not covered by \citet{thomas_slope}). Yellow circles indicate axisymmetric (AS) wave flow, plus signs show data points where weak wave (WW) propagation was observed, upward triangles and squares indicate intermittent bursting (IB) and regular wave flow, respectively.}
\label{regime_flat}
\end{figure}

Here, we present the results of the first series of experiments, which was conducted with flat bottom topography.
Fig.\ref{hm_flat} shows snapshots of four runs performed in the vicinity of the transition boundary along the $\Delta T = 2.5$ K curve, and their respective Hovm\"oller plots. In agreement with the findings of \citet{frueh} and also \citet{thomas_npg}, it is visible that between the axisymmetric (AS) flow regime (Fig.\ref{hm_flat}a) and that of regular propagating wave patterns (Fig.\ref{hm_flat}c and d) lays a certain transitional range, characterized by dispersive weak waves (hereafter denoted with `WW') of fluctuating wave amplitudes (Fig.\ref{hm_flat}b). The diagram of the flat bottom measurements is presented in Fig.\ref{regime_flat}. In the $\Delta T=(7-8)$ K region, data points obtained from earlier experimental studies using the same configuration \citep{torsten,uwe_piv} were also adopted. 

The existence of multiple equilibria (i.e. different flow types at the same values of $Ta$ and $Ro_T$) is apparent from the regime diagram, indicated by the fact that the region `WW' of weak fluctuating waves visibly overlaps with the axisymmetric regime in Fig.\ref{regime_flat}. Occasionally non-unique states emerged as well, in which two given regular wave modes of different wave numbers (typically $m=2$ and $m=3$) were alternating, replacing each other on a typical timescale of $\sim 10$ revolutions. This kind of dynamics is referred to as `intermittent bursting' (IB), and has already been reported by \citet{thomas_npg} (see also references therein). 

The region where the transition from the axisymmetric state to the wave flow regime occurs shows qualitative agreement with the computed neutral linear stability curve (blue graph in Fig.\ref{regime_flat}) obtained by \citet{thomas_slope}. We emphasize, however, that a sharp transition curve cannot be determined for the experiments, since the intermediate transient `WW' state forms a rather smoothly changing continuum between the AS and the steady wave regime.

\subsection{Experiments with sloping bottom endwall}
\begin{figure}[h!]
\noindent\includegraphics[width=8.6cm]{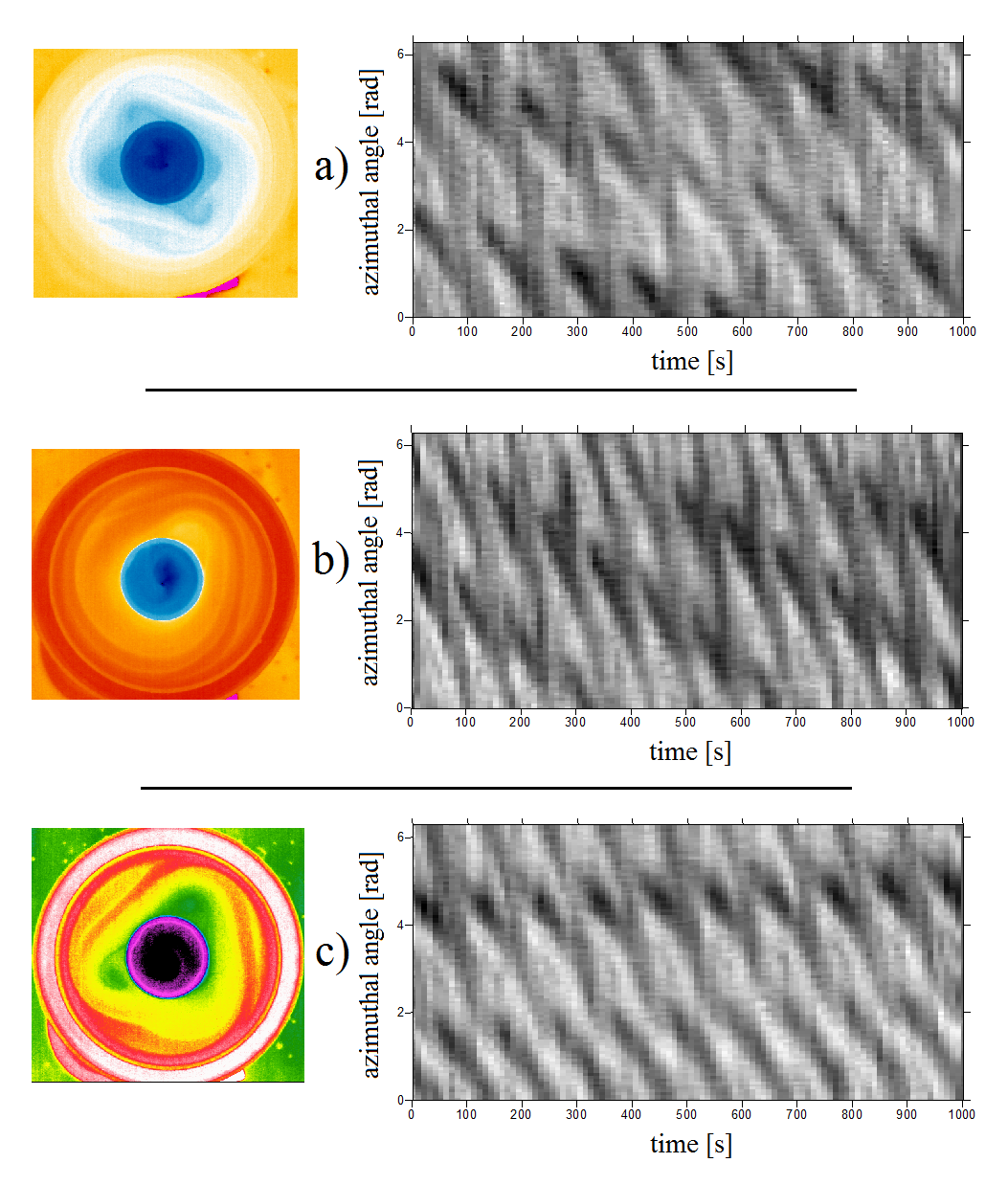}
 \centering
\caption{Thermographic snapshots (left) and Hovm\"oller plots (right) of three experimental runs exhibiting interference vacillation flows with sloping bottom topography. 
a) Involved modes: $m=2,3,5$ ($Ro_T=1.59$; $Ta=6.04\times 10^6$), b) involved modes: $m=3,2,1$ ($Ro_T=1.53$; $Ta=1.35\times 10^7$), c) involved modes $m=2,3,5$ ($Ro_T=1.10$; $Ta=1.84\times 10^7$). Note, that unlike in Fig.\ref{hm_flat}, the contributing wave numbers are not that manifest from pure visual inspection of the plots. The involved modes were determined using the methods described in subsection 4.3.}
\label{hm_slope}
\end{figure}

We now present our findings in the sloping bottom case.
With the introduction of sloping bottom topography, a certain type of \emph{vacillation} was observed which did not appear in the flat bottom case. The steady wave patterns that occur regularly in the flat bottom case (see Hovm\"oller plots in Fig.\ref{hm_flat}c and d) have not been observed in any of our sloping bottom experiments.  
Instead, the dominant baroclinic wave mode appeared in distorted form, modulated periodically by another wave of different
wave number and propagation rate. This superposition phenomenon is illustrated by the snapshots and Hovm\"oller plots of Fig.\ref{hm_slope}. The observed mechanism is referred to as \emph{interference vacillation} (IV), cf. \citet{lindzen}.      

It is generally accepted that the presence of $\beta$-effect enhances dispersion of baroclinic waves, as it was already demonstrated experimentally in a similar set-up (including sloping bottom topography) in the paper by \citet{fultz_kaynor}.

\begin{figure}[h!]
\noindent\includegraphics[width=8.6cm]{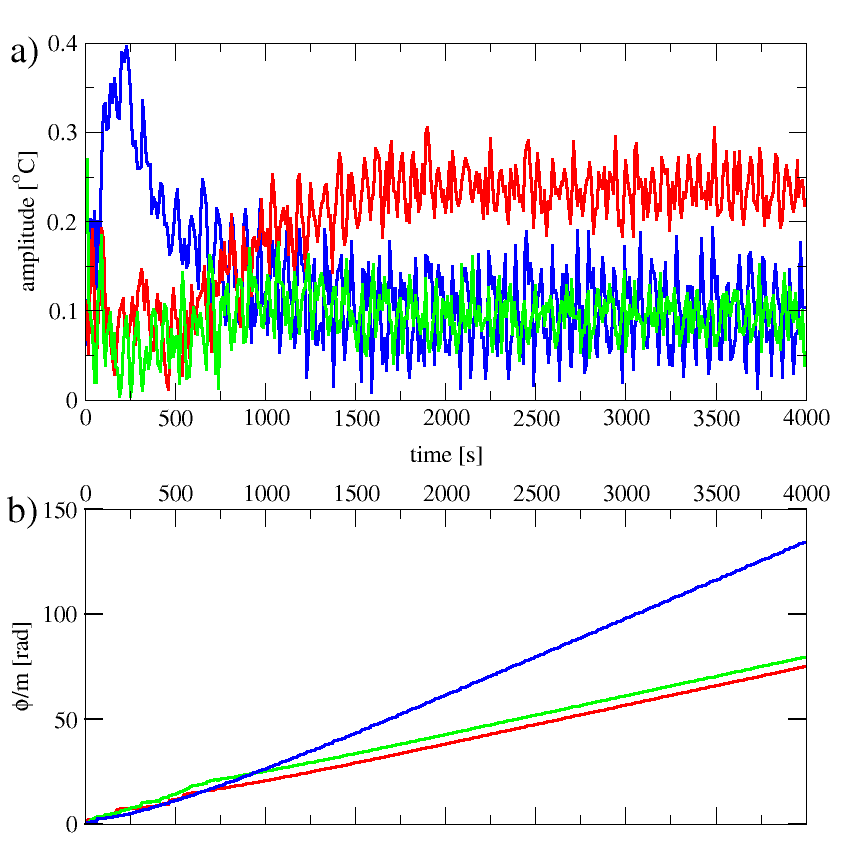}
 \centering
\caption{The temporal development of the amplitudes (a) and `azimuthal distances' of modes $m=3$ (red), $m=4$ (blue) and $m=6$ (green) for an experimental run with sloping bottom ($Ta=2.23\times 10^7$ and $Ro_T=0.93$).}
\label{vac_fourier}
\end{figure}

As an example, a clear signature of dispersion is presented in Fig.\ref{vac_fourier}, for an experiment performed at $Ta=2.23\times 10^7$, $Ro_T=0.93$. Panel a) shows the temporal development of the spatial Fourier amplitudes of components $m=3$, 4 and 6, whereas in panel b) the spatial propagation of these modes is shown (cf. Fig.\ref{fourierdemo}c and d for flat bottom).
As visible in panel a), during the initial transient `spin-up' phase, mode $m=4$ dominated the flow. This component later decayed in amplitude, transferring its energy to the $m=3$ mode and its $m=6$ harmonic (which is the fingerprint of the fine structure of the baroclinic wave, as noted in Section 3). Yet, a perturbation that inherited the four-fold symmetry of the transient initial state did not vanish, but continued to propagate at a significantly larger phase velocity than that of the dominant mode (see panel b). 

It is worth to mention that the phase speed of a given mode generally increased as its amplitude decreased. This is visible from the comparison of Figs.\ref{vac_fourier}a and b: for instance, the propagation of $m=4$ clearly accelerated until $t\approx 700$ s and then stayed constant (see blue curve in panel b), meanwhile its amplitude showed the aforementioned decay until the same $t\approx 700$ s mark, and exhibited stationary behaviour afterwards (see blue curve in panel a). Qualitatively similar amplitude vs. phase speed relations could be observed in the temporal development of the other modes as well. These findings underline the \emph{nonlinear} property of the dynamics present in the baroclinic annulus. 

\begin{figure}[h!]
\noindent\includegraphics[width=8.6cm]{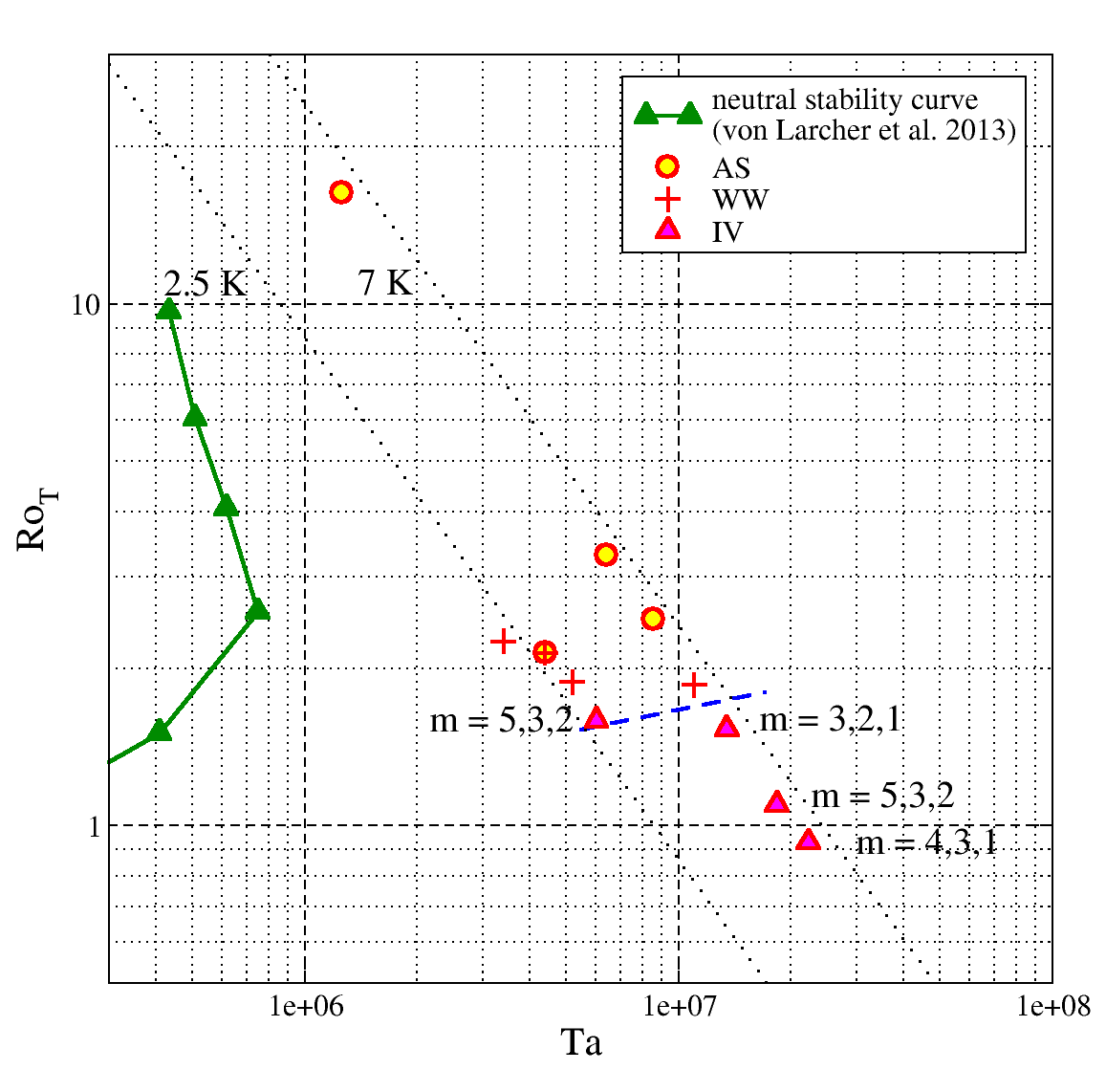}
 \centering
\caption{Regime diagram for the sloping bottom experiments. 
The neutral numerical stability curve for this case is marked by the green solid curve. Circles indicate axisymmetric (AS) wave flow, plus signs show data points where weak wave (WW) propagation was observed, whereas upward triangles indicate interference vacillation (IV) flow. For the data points exhibiting IV, the wave numbers of the interacting wave triads are indicated.}
\label{regime_slope}
\end{figure}

We summarize our findings for the sloping bottom case in the regime diagram of Fig.\ref{regime_slope}.
In these experiments IV wave patterns have been observed in the region of the $Ta-Ro_T$ plane which was characterized by regular steady waves in the flat bottom case.
However, the overall structure of the transition region remained quite similar to the one of Fig.\ref{regime_flat} (the flat bottom neutral stability line is marked by a dashed blue line in Fig.\ref{regime_slope}). This result shows that in this geometry (in contrast with the flat bottom case) the actual regime transition does not occur along the neutral stability curve (green) calculated by \citet{thomas_slope}. The disagreement between the linear stability analysis and the actual flow (as well as the presence of IV) implies \emph{nonlinear} wave-wave and wave-background flow interactions that are enhanced by the sloping bottom.
As visible in Fig.\ref{regime_slope}, the neutral linear stability curve exhibits a sharp reversal at higher values of $Ro_T$, as well as a general destabilizing effect of the sloping bottom end wall, which manifests itself in the fact that the curve of the sloping case envelopes that of the flat bottom case (see also Fig.\ref{thomas_replot}). Thus, without nonlinear interactions, regular wave propagation would be observed all over the experimentally investigated parameter region in the sloping bottom configuration.
       
\subsection{Mode identification in the IV regime}
The simplest form of wave-wave interactions is a three-wave resonance, in which two coexisting waves of different wave numbers and drift rates excite and amplify a third mode \citep[e.g.][]{plumb}. The selection criteria of such resonant \emph{triads} are that the wave vectors $\mathbf{k}_i$ of the modes involved in should add up to zero, as
\begin{equation}
\mathbf{k}_m\pm\mathbf{k}_{m'}\pm\mathbf{k}_{m''}=0,
\label{wave_vectors}
\end{equation}   
and also the following relationship should hold between the spatial phases $\phi_i$ of the three interacting modes \citep{frueh}:
\begin{equation}
\phi_m(t)-\phi_{m'}(t)-\phi_{m''}(t)\equiv \varphi_{m-m'-m''}(t)\approx \rm{const.}
\label{drifts}
\end{equation} 
For an exactly resonant situation, termed `phase lock', the time series $\varphi_{m-m'-m''}(t)$ would give a constant value between 0 and $2 \pi$.
One can therefore define a probability density function $\rho_{m-m'-m''}(\phi)$ of $\varphi_{m-m'-m''}(t)$, that practically means a histogram of the time series. If marked phase coherence (i.e. triad resonance) is present, then $\rho_{m-m'-m''}(\phi)$ should exhibit a sharp peak, whereas, if (\ref{drifts}) was not satisfied, the distribution would be rather flat.

\begin{figure}[h!]
\noindent\includegraphics[width=8.6cm]{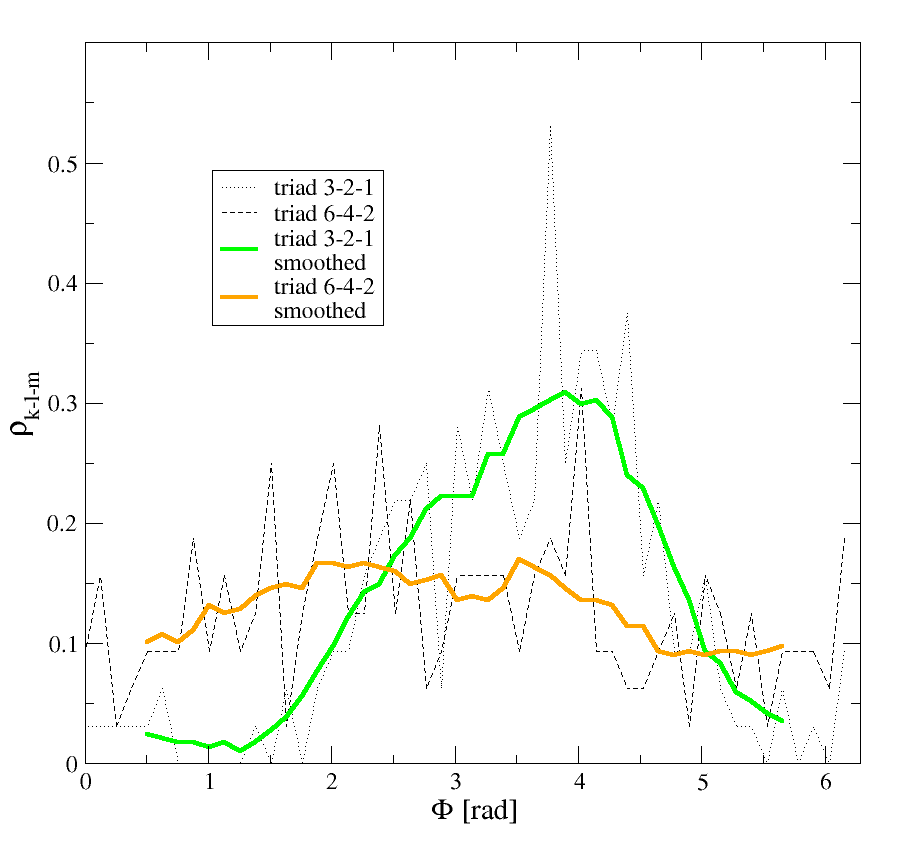}
 \centering
\caption{Triad locking probality density functions for triads $m=3,2,1$ and $m=6,4,2$, and their (9-point) moving-averaged, `smoothed' versions in a sloping bottom experiment. See also the legend. ($Ro_T=1.53$; $Ta=1.35\times 10^7$)}
\label{phaselock}
\end{figure}

In terms of the \emph{azimuthal} wave numbers, studied throughout this paper, (\ref{wave_vectors}) can simply be formulated as:
\begin{equation}
m\pm m' \pm m'' =0.
\label{azimuthal}
\end{equation}
As mentioned in Section 3, our Fourier transform scheme was based on components of integer wave numbers $m=1$ to 6. Among these, five triad combinations can be selected that fulfil (\ref{azimuthal}).
In order to identify the most dominant interacting waves, the density functions $\rho_{m-m'-m''}(\phi)$ of each triad were analysed for all four experiments where IV was detected.  
Two examples of $\rho_{m-m'-m''}(\phi)$ are shown in Fig.\ref{phaselock}. In this particular case $\rho_{3-2-1}(\phi)$ exhibits a well-defined peak, whereas  $\rho_{6-4-2}(\phi)$ does not show any significant coherence (see also the corresponding Hovm\"oller plot and the parameters in Fig.\ref{hm_slope}b).

The analysis of the triad locking density functions combined with that of the amplitude and phase time series (e.g. the ones in Fig.\ref{vac_fourier}) enabled us to identify the most dominantly interacting modes in all four experiments where IV was detected. The resulting wave numbers of the dominant triads are marked in the regime diagram of Fig.\ref{regime_slope}.

\begin{figure}[h!]
\noindent\includegraphics[width=8.6cm]{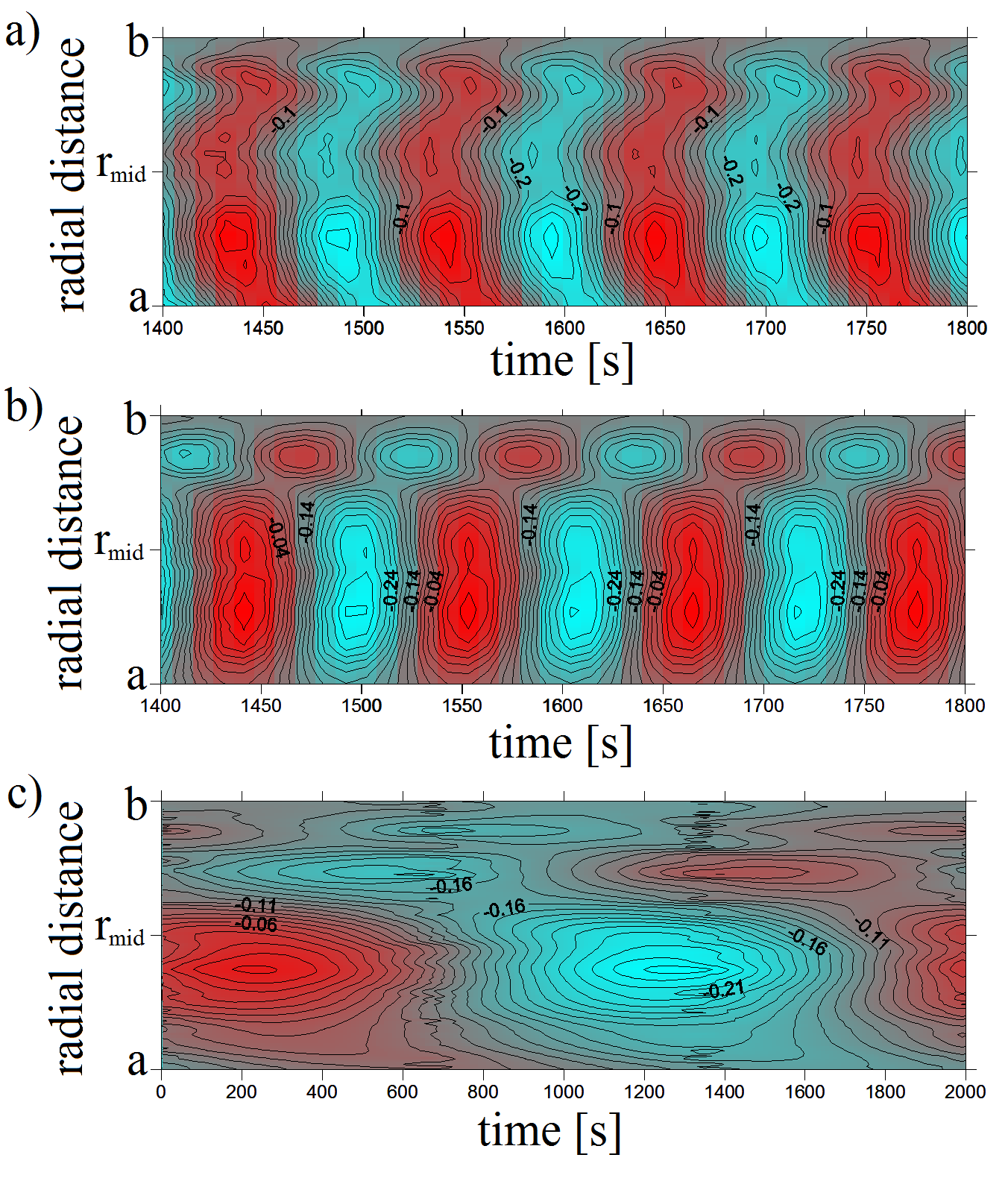}
 \centering
\caption{Hovm\"oller plots of the filtered \emph{radial} patterns that correspond to the frequencies of the azimuthal wave motions the triad of $m=3$ (a), $m=5$ (b) and $m=2$ (c). Red colours denote regions of higher temperatures. The contour spacing was set to 0.1 K. Note the wider temporal range in panel c). ($Ro_T=1.099$; $Ta=1.83\times 10^7$)}
\label{radial_hm}
\end{figure}

Although this investigation focused on the azimuthal wave numbers, it is to be noted that the relation (\ref{azimuthal}) is a necessary but per se not sufficient condition to fulfil the selection criterion (\ref{wave_vectors}) of the wave vectors. Equations analogous to (\ref{azimuthal}) exist for the radial ($l$) and vertical ($n$) components of the triad-forming wave vectors, in the form of:
\begin{equation}
l\pm l' \pm l'' =n\pm n' \pm n'' =0.
\label{radial}
\end{equation}
Since the vertical behaviour could not be directly detected with our measurement techniques, we analysed the \emph{radial} patterns corresponding to the interacting azimuthal modes. To reveal these structures, path-wise temperature series were extracted from each thermographic frame (as described in detail in Section 3), this time along a radial section ranging from $a$ to $b$ and consisting of 110 grid points at each time instant. As the next step, bandpass filtering has been applied to all of these 110 temperature records in the (disjoint) $f_m = m \cdot c_m  / (2\pi)\pm 1.5\times 10^{-4}\mathrm{1/s}$ frequency ranges, taken with the angular velocities $c_m$ of the azimuthal modes $m$, obtained from (\ref{omega}). Then the filtered time series were recombined into Hovm\"oller plots like the ones presented in Fig.\ref{radial_hm} as typical examples. These show the radial wave motion associated with the frequencies of the dominant interacting triad ($m=5, 3$ and $2$) of the experimental run at $Ro_T=1.099$ and $Ta=1.83\times 10^7$. 

Visibly, the patterns in panel a) ($m=3$) and b) ($m=5$) are more complex than pure standing wave modes characterized by a single value of $l$, yet a regular, spatially correlated structure is present. In panel b) even a steady node can be observed at radial distance $r \approx (r_{\rm mid}+b)/2$ which separates an inner domain from an outer one, oscillating at a clear phase lag of one-fourth of a period. These Hovm\"oller plots indicate that even in these narrow frequency bands $f_m$ the radial dynamics is driven by the superposition of at least two standing wave modes, some of which fulfil (\ref{radial}), while others do not.

\subsection{Time-reversal asymmetry and traces of nonlinearity}
\begin{figure}[h!]
\noindent\includegraphics[width=7cm]{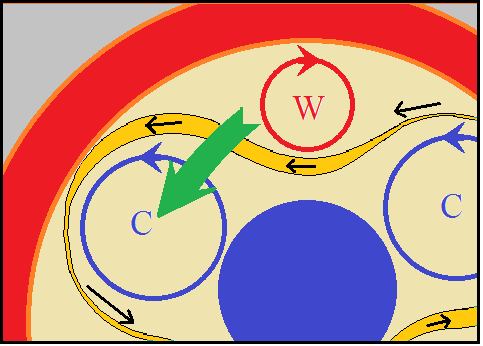}
 \centering
\caption{Schematic draft of the spatial asymmetries between the front and the wake sides of a baroclinic lobe. Note that the jet (orange), meandering between cold (C) and warm (W) vorices accelerates (see arrows) and therefore compresses at the front side, compared to the wake side.}
\label{jet}
\end{figure}

It has been generally observed that in the regular, non-dispersive baroclinic waves of the flat bottom configuration (e.g. the one in Fig.\ref{hm_flat}d) a jet is formed along the surface streamlines, encompassing the `cold' eddies (see orange strip in the schematic drawing of Fig.\ref{jet}). As the wave pattern drifts (green arrow), the streamlines at the front of the cold lobes compress (the jet accelerates) and thus the temperature gradient steepens (``nonlinear steepening''). Meanwhile, in the wake of the cold vortices the gradient flattens. This spatial asymmetry, a sign of nonlinear wave propagation, manifests itself as the time reversal asymmetry of the temperature time series obtained from co-rotating temperature measurements. A similar asymmetry has been reported by \citet{gyure} in the regime of \emph{geostrophic turbulence} as well, in experiments with flat bottom end wall. In the sloping bottom case, however, this structure is expected to be distorted by the wave-wave interactions. The modulation caused by the other wave mode(s) involved in the interference vacillation destabilizes the jet and leads to mixing, which may reduce the asymmetry, thus providing further evidence for the existence of wave-wave interactions. This hypothesis has been tested by analysing point-wise surface temperature records from our experiments for both geometries. 

The time series were extracted from the same raw thermographic images as the azimuthal temperature field data, after selecting a location at radial distance $r_{\rm mid}\equiv (a+b)/2$. Naturally, the sampling time was the same as mentioned in Section 3, i.e. $\Delta t = 10$ s. In case of data loss (which may occur, e.g., because of the occasional automatic recalibration of the microbolometers in the infrared camera), linear interpolation was applied to yield equally spaced data.    
 
The aforementioned temporal asymmetry has been quantified in terms of two simple quantities by \citet{gyure}:
the ratio of the \emph{number} of warming and cooling steps $N_w/N_c$ in the time series of a given experimental run, and the ratio of their mean \emph{magnitudes}, averaged over the record, $\langle \Delta T_w \rangle / \langle \Delta T_c \rangle$.
The significant deviation of these parameters from unity (i.e. the presence of temporal asymmetry in \emph{stationary} time series) can be interpreted as a sign of nonlinearity, as mentioned above. The same type of analysis has been repeated in the present study.

\begin{figure}[b!]
\noindent\includegraphics[width=8.6cm]{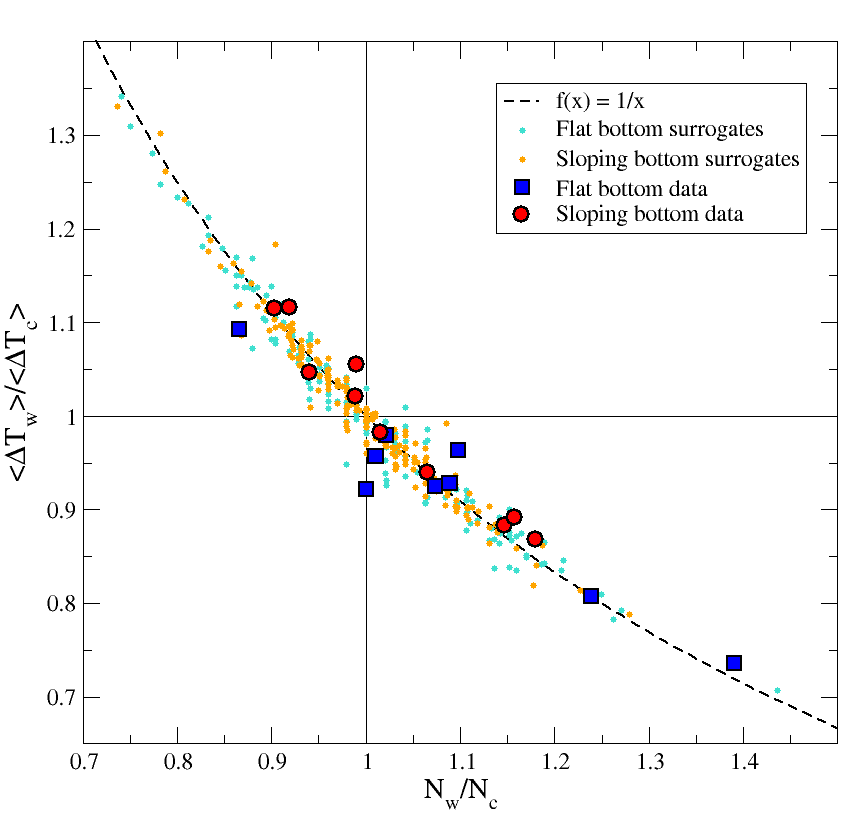}
 \centering
\caption{Correlation plot between step number ratios and average step size ratios from surface temperature records for both bottom end wall configurations. The values corresponding to the `shuffled' surrogate time series are also presented. Dashed line denotes $1/x$. See also the legend.}
\label{asymm}
\end{figure}

Similarly to \citet{gyure}, to define a 95\% confidence interval, 20 random shuffled time series were produced from the original signals of the experimental runs, for each record. In order to achieve this, the iterative Fourier surrogate method by \citet{schreiber}, and the open-source software package TiSeAn 3.0.1 (by the same authors) was applied. Besides conserving the histogram, this procedure also keeps Fourier amplitudes intact (by shuffling only the phases of the spectral components), but destroys all nonlinear correlations in the signal. The surrogate data produced this way should theoretically exhibit full symmetry (i.e. both $N_w/N_c=\langle \Delta T_w \rangle / \langle \Delta T_c \rangle\equiv 1$), yet, due to the finite finite length of the time series, certain deviations are present. However, the \emph{distribution} of the parameter values for these shuffled signals is symmetric to 1, as shown by the small turquoise and orange dots in the scatter plot of Fig.\ref{asymm}, representing the values obtained for the flat and sloping bottom surrogates, respectively. 

If the conditions of stationarity are fulfilled, the above defined two measures of asymmetry should be inversely proportional to each other; the smaller is the number of steps in one direction (compared to that of the other direction), the larger the average relative size of the magnitude of these steps should be, to avoid increasing or decreasing trends.
Thus, the fact that all the data points of Fig.\ref{asymm} -- both from surrogates and actual signals -- are scattered along the $f(x)=1/x$ curve (dashed line) implies that stationarity holds for our experimental data. 

The most important observation is that the scattering of the data points obtained from the experiments with flat bottom end wall (blue squares in Fig.\ref{asymm}) shows a marked deviation from unity (and from the distribution of the surrogates), and is clearly asymmetric. For the vast majority of the realizations $N_w/N_c>1$ and $\langle \Delta T_w \rangle / \langle \Delta T_c \rangle<1$ can be found, indicating nonlinearity. We note, that the extent and direction of this asymmetry obtained close to the axisymmetric regime is in good agreement with the findings of \citet{gyure}, although, as mentioned above, their experiments were performed in the turbulent regime (at $Ta=9.43\times 10^9$ and $Ro_T=0.035$).  
As expected, for the sloping bottom experiments (red circles in Fig.\ref{asymm}) no such tendency was observed; the distribution of these data points is fairly symmetric with respect to 1 and overlaps with that of their surrogates. 
This result confirms our assumption that wave-wave interactions in the sloping bottom experiments generally destroy the structure of the jet encompassing the cold lobes and thus lead to the reduction of steepening and to the homogenization of the flow field.    
\section{Discussion}  
Two series of laboratory experiments have been performed in a differentially heated rotating annulus set-up, recorded by a co-rotating high precision infrared camera and evaluated using digital image processing techniques, spatial Fourier transforms and standard tools of nonlinear time series analysis.
The most important motivation for these measurements was to compare the alignment of the boundary between the axisymmetric and the regular wave flow regimes on the parameter space spanned by the Taylor and thermal Rossby numbers with the curve of neutral linear stability, obtained in a numerical study by \citet{thomas_slope}. 

In the first series of experiments the bottom end wall of the tank was flat. As far as this geometry is concerned, our experimental results showed a qualitative agreement with the numerical results. It was pointed out, that between the axisymmetric state and the domain of robust regular baroclinic waves a transient behaviour can be observed, characterized by uncertain, fluctuating wave patterns. The occurrence of such transient flows makes it difficult to define a sharp transition from the axisymmetric to the wave state. This explains why we find qualitative agreement only. The transient regime exhibits multiple equilibria: vacillating decaying waves and axisymmetric flow can both exist at the same values of $Ta$ and $Ro_T$.
Because of their short lifespan and fluctuating amplitudes, no conclusive dispersion relation of these waves could be obtained, but from the visual inspection of Hovm\"oller plots as the one in Fig.\ref{hm_flat}b, it can certainly be stated that the drift rates of these waves clearly depend (among other things) on the wave number, i.e. dispersion is present. Once the stable wave flow regime is reached, dispersion ceases to exist; all the (physically meaningful) Fourier components propagate at the same rate, maintaining the complex pattern of the steady baroclinic waves.

\begin{figure}[b!]
\noindent\includegraphics[width=8.6cm]{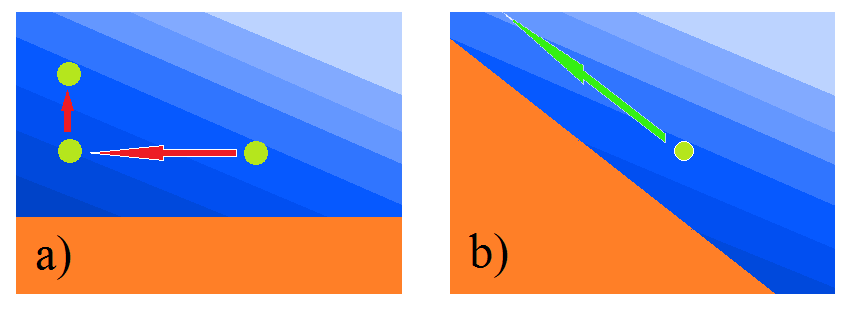}
 \centering
\caption{Schematic drawing of the qualitative background of baroclinic instability. For a flat bottom topography (a) a small horizontal displacement of a fluid parcel leads to a larger displacement, i.e. instability occurs. If the bottom topography is steep enough (b), geometric constraints inhibit destabilization in the bottom region.}
\label{moricka}
\end{figure}

The second series of measurements was performed with a sloping bottom obstacle placed in the annulus. The key problem that has been addressed in this case was whether this modification stabilizes or destabilizes the system.
According to the naive reasoning, often used in the literature (see, e.g. \citet{realvallis}), a slope that is steep enough would generally \emph{stabilize} the flow. The schematic drawing of Fig.\ref{moricka} illustrates this argument. In the `standard' case of baroclinic instability, a fluid parcel which is moved by some initial perturbation parallel to the (flat) bottom, though travelling along an equipotential, may reach a region of higher density, since the thermal wind balance keeps the isotherms tilted (Fig.\ref{moricka}a). From here, the buoyancy force lifts the parcel until it reaches its original isotherm -- at a \emph{larger} geopotential. The increased potential energy of the system is responsible for the excitation of baroclinic waves. If, however, the slope of the bottom is steeper than that of the isothermal surfaces, a parcel moving parallel to the bottom is pushed back to its initial position by the buoyancy, therefore no instability can arise in the bottom region (Fig.\ref{moricka}b). 

Within the framework of the geostrophic theory of thermal wind, one can easily obtain an order-of-magnitude estimate of the slope $s_T$ of isotherms (see, e.g. \citet{vallis}) in the form of
\begin{equation}
s_T=\frac{f\cdot U}{g\alpha\Delta T},
\label{gamma_t}
\end{equation}
where $U$ denotes the characteristic velocity of the zonal background flow (the other symbols denote the same quantities as in (\ref{Ro})). Fortunately, there exist earlier PIV (Particle Image Velocimetry) based measurement data on the velocity $U$ from the same $Ta-Ro_T$ parameter range and the same geometry -- albeit for flat bottom only -- as reported by \citet{uwe_piv}. Based on these data, we can take $U\sim 10^{-3}$ m/s as a safe (over-) estimation of the background flow. In the same spirit, we can substitute the smallest applied temperature difference $\Delta T=2.45$ K, as well as the largest Coriolis parameter $f=2\Omega=1.12$ s$^{-1}$ into (\ref{gamma_t}), which yields $s_T^{\rm max}\approx 0.2$, still significantly smaller than the slope $\tan(\gamma)= 0.7$ of our bottom topography. Therefore, based on this argument, one would expect that the instability is inhibited by the slope, and thus the wave flow regime in the $Ta-Ro_T$ plane would shrink compared to the flat bottom case.

\begin{figure}[h!]
\noindent\includegraphics[width=7cm]{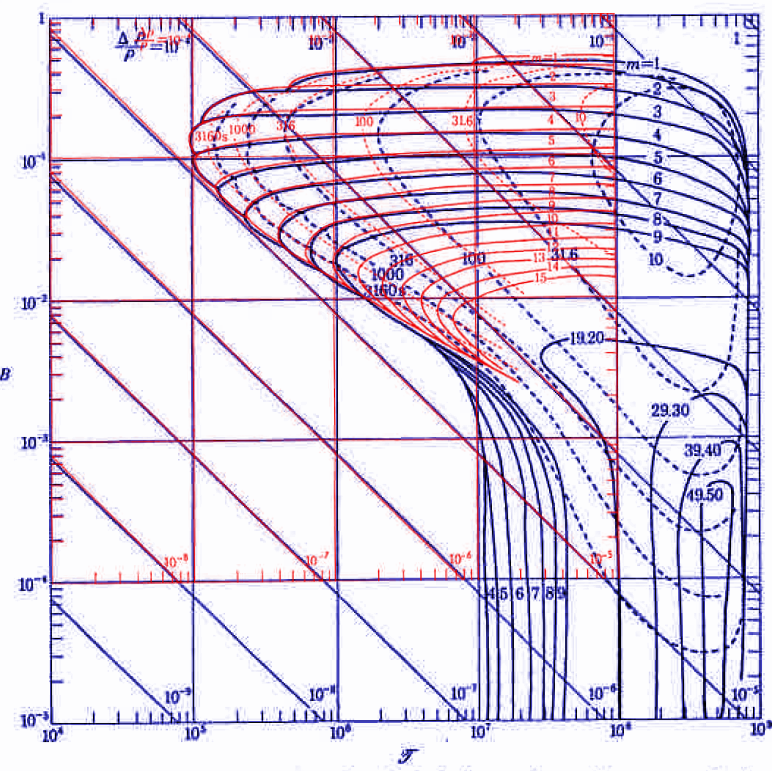}
 \centering
\caption{Regime diagrams obtained by \citet{mason} for flat (red) and gently sloping (blue) bottom end walls, c.f. Fig.\ref{thomas_replot}.}
\label{mason_plots}
\end{figure}

A more advanced theoretical approach has been followed by \citet{mason}, based on an extension of the classic baroclinic instability theory of \citet{eady} to include Ekman layers. Mason obtained regime diagrams for different top and bottom endwall configurations, including ones similar to the cases studied here. In Fig.\ref{mason_plots} we repeat Figs.20 and 21 of this original paper, overlaid onto each other to highlight the differences. Note, that the geometric properties of the annulus described there differ from those of the one used in our study, and so are the control parameters. Yet, parameter $\cal{T}$ of the horizontal axis is equivalent to $Ta$ and $B$ of the vertical axis is proportional to $Ro_T$, thus these curves are suitable for qualitative comparison with Fig.\ref{thomas_replot}. 
The main distinctive feature of the regime diagram for the sloping case (blue graph in Fig.\ref{mason_plots}) compared to that of the flat bottom case (red contour) is that the `lower axisymmetric' regime has been destabilized (in contrast with the previous naive reasoning). On the other hand, careful inspection reveals a slight tendency for stabilization around the boundary between the wave flow regime and the `upper' axisymmetric region: the boundary shifted towards lower values of $B$ with the introduction of sloping bottom.   
It is also to be noted that the modified Eady model, being a linear theory, is confined to the limit where $\tan(\gamma)$ is small (and also smaller than the prescribed slope $s_T=0.65$ of the isotherms used in their calculations). 

To determine the regime boundaries of Fig.\ref{thomas_replot}, \citet{thomas_slope} have carried out a \emph{linear} stability analysis based on the spectral decomposition of the azimuthal modes, and thus, obviously could not resolve wave-wave interactions. However, as it is visible from our experimental data, the actual physical eigenmodes of the system are far not sinusoidal; single normal modes cannot be found but they usually show up in pairs or triplets. As an example, we refer to Figs.\ref{fourierdemo}d and \ref{vac_fourier}b, which demonstrate that the spectral `fingerprint' of the three-fold symmetric eigenmode always contains the harmonic $m=6$, too, and a pattern that corresponds to a single $m=3$ Fourier mode \emph{per se} simply cannot be excited. When the slope is present, this difference becomes critical, since in this configuration multiple physical eigenmodes can coexist that propagate at different drift rates, giving way to energy transfer between modes. Such wave-wave interactions can either destabilize or stabilize the system; latter may be responsible for the disagreement between the numerical and experimental results.      

The fact that sloping bottom topography leads to dispersive propagation of co-existing waves in a differentially heated rotating annulus was first noticed by \citet{fultz_kaynor}. By that time if was already well known, that in a rotating set-up over such a bottom end wall barotropic \emph{topographic Rossby waves} may be excited (independently of the thermal boundary conditions, driven by the conservation of potential vorticity only). Therefore, it has been naturally assumed by the authors that such waves are responsible for the observed behaviour as they modulate the waves of baroclinic instability. Because of the limitations of their early measurement techniques, the results of this study are not exactly conclusive, yet they triggered further study of wave dispersion in similar set-ups. \citet{pfeffer} and \citet{uwe_piv} reported dispersive phenomena in the vicinity of the regime boundary between the `upper' axisymmetric and the regular wave flow regime and betwen two wave regimes, respectively, studying a flat bottom set-up with \emph{free surface}. Occasionally, the very same apparatus that is investigated in the present paper -- without slope but with free surface -- also exhibited IV in the regime boundary region (besides the IB-type behaviour mentioned in subsection 4.1), as noted by \citet{thomas_npg} and \citet{uwe_piv}.

In our set-up, a $\beta$ parameter can be defined as follows \citep{vallis}:
\begin{equation}
\beta=\frac{f\cdot\tan(\gamma)}{d_{\rm mid}},
\label{beta}
\end{equation}  
with $d_{\rm mid}=10.85$ cm being the water depth at mid radius $r_{\rm mid}$.
In a two-layer model \citet{mansbridge} investigated the effect of different values of $\beta$ on the flow properties, and found that neighbouring modes can coexist if $\delta\equiv\beta Ro_T / \sqrt{Ek}$ is larger than a certain threshold (the value of which depends on mode number $m$, and was found to be between 10 and 63). Here the Ekman number $Ek$ can be defined as: $Ek=\nu/(\Omega L^2)$, where $\nu\approx 10^{-6}$ m$^2$/s is the kinematic viscosity of water and the characteristic length was taken to be $L=r_{\rm mid}=(a+b)/2$. The values of $\delta$ in our sloping bottom runs were ranging from 340 to 950, i.e. way over the aforementioned theoretical threshold, consistently with our detection of IV.  
  
Further experimental studies are under consideration, involving direct velocity measurements applying PIV technique to
gain information on the flow field in the set-up. Using that data one could   
separate the propagation rate of the different baroclinic wave modes from the drift originating from the background flow.
This would make possible to obtain dispersion relations and -- in the case of strongly nonlinear patterns -- amplitude-velocity relations which would lead to a deeper understanding of baroclinic waves in general, and would support a better comparison of the results to the predictions of different theories. We note that, assuming quasi-geostrophic connection between the time-averaged radial gradient of the observed surface temperature field and the zonal background flow, we already attempted to distinguish wave propagation from the flow. These trials however did not lead to any conclusive connection between the aforementioned quantities, which can be explained by the presence of multiple equilibria (i.e. that similar surface patterns may correspond to different flow states if the whole three-dimensional water body is concerned). Therefore PIV-based velocity measurements at different vertical levels would be crucial, as they may shed light on these issues as well. 

\section*{Acknowledgements}
The authors thank Yongtai Wang for his crucial help in performing the measurements. The fruitful discussions with Wolf-Gerrit Fr\"uh and Kiril Alexandrov are also highly acknowledged. This work has been funded by the German Science Foundation (DFG)
and is part of the DFG priority program MetStr\"om (SPP 1276).

\end{document}